\documentclass[aps,prl,twocolumn,showpacs,floatfix,superscriptaddress]{revtex4-1}

  \usepackage[utf8]{inputenc}
  \usepackage[T1]{fontenc}
  \usepackage{float} 
  \usepackage{color}  

\usepackage{graphicx}
\usepackage{amsmath}
\usepackage{amssymb}
\usepackage{bbm}
\usepackage{indentfirst}
\usepackage{dcolumn}
\usepackage{soul}
\usepackage{mathrsfs}
\usepackage{bigints}
\usepackage{tikz}
\usetikzlibrary{calc}
\usetikzlibrary{arrows.meta}

\begin{document}

\title{Effective damping enhancement in noncollinear spin structures}

\author{Levente R\'{o}zsa}
\email{rozsa.levente@physnet.uni-hamburg.de}
\author{Julian Hagemeister}
\author{Elena Y. Vedmedenko}
\author{Roland Wiesendanger}
\affiliation{Department of Physics, University of Hamburg, D-20355 Hamburg, Germany}
\date{\today}

\begin{abstract}

Damping mechanisms in magnetic systems determine the lifetime, diffusion and transport properties of magnons, domain walls, magnetic vortices, and skyrmions. 
Based on the phenomenological Landau--Lifshitz--Gilbert equation, here the effective damping parameter in noncollinear magnetic systems is determined describing the linewidth in resonance experiments or the decay parameter in time-resolved measurements. It is shown how the effective damping can be calculated from the elliptic polarization of magnons, arising due to the 
noncollinear spin arrangement. It is concluded that the effective damping is larger than the Gilbert damping, 
and it may significantly differ between excitation modes. Numerical results for the effective damping are presented for the localized magnons in isolated skyrmions, with parameters based on the Pd/Fe/Ir(111) model-type system.

\end{abstract}

\maketitle

Spin waves (SW) or magnons as elementary excitations of magnetically ordered materials have attracted significant research attention lately. The field of magnonics\cite{Kruglyak} concerns the creation, propagation and dissipation of SWs in nanostructured magnetic materials, where the dispersion relations can be adjusted by the system geometry. A possible alternative for engineering the properties of magnons is offered by 
noncollinear (NC) spin structures\cite{Garst} 
instead of collinear ferro- (FM) or antiferromagnets (AFM). SWs are envisaged to act as information carriers, where one can take advantage of their low wavelengths compared to electromagnetic waves possessing similar frequencies\cite{Schwarze}. Increasing the lifetime and the stability of magnons, primarily determined by the relaxation processes, is of crucial importance in such applications.

The Landau--Lifshitz--Gilbert (LLG) equation\cite{Landau,Gilbert} is commonly applied for the quasiclassical description of SWs, where relaxation is encapsulated in the dimensionless Gilbert damping (GD) parameter $\alpha$. The lifetime of excitations can be identified with the resonance linewidth in frequency-domain measurements such as ferromagnetic resonance (FMR)\cite{Satywali}, Brillouin light scattering (BLS)\cite{Serga} or broadband microwave response\cite{Onose}, and with the decay speed of the oscillations in time-resolved (TR) experiments including magneto-optical Kerr effect microscopy (TR-MOKE)\cite{Liu} and scanning transmission x-ray microscopy (TR-STXM)\cite{Finizio}. Since the linewidth is known to be proportional to the frequency of the magnon, measuring the ratio of these quantities is a widely applied method for determining the GD in FMs\cite{Schwarze,Satywali}. An advantage of AFMs in magnonics applications\cite{Abraha,Camley} is their significantly enhanced SW frequencies due to the exchange interactions, typically in the THz regime, compared to FMs with GHz frequency excitations. However, it is known that the linewidth in AFM resonance is typically very wide 
because it scales with a larger effective damping parameter $\alpha_{\textrm{eff}}$ than the GD $\alpha$\cite{Gurevich}.

The tuning of the GD can be achieved in magnonic crystals by combining materials with different values of $\alpha$. It was demonstrated in Refs.~\cite{Kruglyak2,Tiwari,RomeroVivas} that this leads to a strongly frequency- and band-dependent $\alpha_{\textrm{eff}}$, based on the relative weights of the magnon wave functions in the different materials.

Magnetic vortices are two-dimensional NC spin configurations in easy-plane FMs with an out-of-plane magnetized core, 
constrained by nanostructuring them in dot- or pillar-shaped magnetic samples. The excitation modes of vortices, particularly their translational and gyrotropic modes, have been investigated using collective-coordinate models\cite{Guslienko} based on the Thiele equation\cite{Thiele}, linearized SW dynamics\cite{Ivanov,Mruczkiewicz}, numerical simulations\cite{Kruger} and experimental techniques\cite{Novosad,Guslienko2,Kamionka}. It was demonstrated theoretically in Ref.~\cite{Kruger} that the rotational motion of a rigid vortex excited by spin-polarized current displays a larger $\alpha_{\textrm{eff}}$ than the GD; a similar result was obtained based on calculating the energy dissipation\cite{Khvalkovskiy}. However, due to the unbounded size of vortices, the frequencies as well as the relaxation rates sensitively depend on the sample preparation, particularly because they are governed by the magnetostatic dipolar interaction.

In magnetic skyrmions\cite{Bogdanov}, the magnetic moment directions wrap the whole unit sphere. In contrast to vortices, isolated skyrmions need not be confined for stabilization,
and are generally less susceptible to demagnetization effects\cite{Kiselev,Schwarze}. The SW excitations of the skyrmion lattice phase have been investigated theoretically\cite{Petrova,Zang,Mochizuki} and subsequently measured in bulk systems\cite{Schwarze,Onose,Ehlers}. It was calculated recently\cite{dosSantos} that the magnon resonances measured via electron scattering in the skyrmion lattice phase should broaden due to the NC structure. Calculations predicted the presence of different localized modes concentrated on the skyrmion for isolated skyrmions on a collinear background magnetization\cite{Lin,Schutte,Kravchuk} and for skyrmions in confined geometries\cite{Mruczkiewicz,Kim,Beg}. From the experimental side, the motion of magnetic bubbles in a nanodisk was investigated in Ref.~\cite{Buttner}, and it was proposed recently that the gyration frequencies measured in Ir/Fe/Co/Pt multilayer films is characteristic of a dilute array of isolated skyrmions rather than a well-ordered skyrmion lattice\cite{Satywali}. However, the lifetime of magnons in skyrmionic systems based on the LLG equation is apparently less explored.

It is known that NC spin structures may influence the GD via emergent electromagnetic fields\cite{Zang,Bisig,Akosa2} or via the modified electronic structure\cite{Akosa,Freimuth}. 
Besides determining the SW relaxation process, the GD also plays a crucial role in the motion of domain walls\cite{Schlickeiser,Selzer,Yuan} and skyrmions\cite{Jiang,Lin2,Ritzmann} driven by electric or thermal gradients, both in the Thiele equation where the skyrmions are assumed to be rigid and when internal deformations of the structure are considered. Finally, damping and deformations are also closely connected to the switching mechanisms of superparamagnetic particles\cite{Brown,Coffey} and vortices\cite{Martens}, as well as the lifetime of skyrmions\cite{Hagemeister,Rozsa,Bessarab}.

The $\alpha_{\textrm{eff}}$ in FMs depends on the sample geometry due to the shape anisotropy\cite{Gurevich,Kalarickal,Kambersky}. It was demonstrated in Ref.~\cite{Kambersky} that $\alpha_{\textrm{eff}}$ is determined by 
a factor describing the ellipticity of the magnon polarization caused by the shape anisotropy. Elliptic precession and GD were also investigated by considering the excitations of magnetic adatoms on a nonmagnetic substrate\cite{Guimaraes}. The calculation of the eigenmodes in 
NC systems, e.g. in Refs.~\cite{Satywali,Mruczkiewicz,Kravchuk}, also enables the evaluation of the ellipticity of magnons, but this property apparently has not been connected to the damping so far.

Although different theoretical methods for calculating $\alpha_{\textrm{eff}}$ have been applied to various systems, a general description applicable to all NC structures seems to be lacking. Here it is demonstrated within a phenomenological description of the linearized LLG equation how magnons in NC spin structures relax with a higher effective damping parameter $\alpha_{\textrm{eff}}$ than the GD. 
A connection between $\alpha_{\textrm{eff}}$ and the ellipticity of magnon polarization forced by the NC spin arrangement is established. The method is illustrated by calculating the excitation frequencies of isolated skyrmions, considering experimentally determined material parameters for the Pd/Fe/Ir(111) model system\cite{Romming}. 
It is demonstrated that the different localized modes display different effective damping parameters, with the breathing mode possessing the highest one.

The LLG equation reads
\begin{eqnarray}
\partial_{t}\boldsymbol{S}=-\gamma'\boldsymbol{S}\times\boldsymbol{B}^{\textrm{eff}}-\alpha\gamma'\boldsymbol{S}\times\left(\boldsymbol{S}\times\boldsymbol{B}^{\textrm{eff}}\right), \label{eqn1}
\end{eqnarray}
with $\boldsymbol{S}=\boldsymbol{S}\left(\boldsymbol{r}\right)$ the unit-length vector field describing the spin directions in the system, $\alpha$ the GD and $\gamma'=\frac{1}{1+\alpha^{2}}\frac{ge}{2m}$ the modified gyromagnetic ratio (with $g$ being the $g$-factor of the electrons, $e$ the elementary charge and $m$ the electron mass). Equation~(\ref{eqn1}) describes the time evolution of the spins governed by the effective magnetic field $\boldsymbol{B}^{\textrm{eff}}=-\frac{1}{\mathcal{M}}\frac{\delta \mathcal{H}}{\delta\boldsymbol{S}}$, with $\mathcal{H}$ the Hamiltonian or free energy of the system in the continuum description and $\mathcal{M}$ the saturation magnetization.

The spins will follow a damped precession relaxing to a local minimum $\boldsymbol{S}_{0}$ of $\mathcal{H}$, given by the condition $\boldsymbol{S}_{0}\times\boldsymbol{B}^{\textrm{eff}}=\boldsymbol{0}$. 
Note that generally the Hamiltonian represents a rugged landscape with several local energy minima, corresponding to e.g. FM, spin spiral and skyrmion lattice phases, or single objects such as vortices or isolated skyrmions. The excitations can be determined by switching to a local coordinate system\cite{Schutte,Mruczkiewicz,Lin2} with the spins along the $z$ direction in the local minimum, $\tilde{\boldsymbol{S}}_{0}=\left(0,0,1\right)$, and expanding the Hamiltonian in the variables $\beta^{\pm}=\tilde{S}^{x}\pm\textrm{i}\tilde{S}^{y}$, introduced analogously to spin raising and lowering or bosonic creation and annihilation operators in the quantum mechanical description of magnons\cite{Holstein,Dyson,Maleev}. The lowest-order approximation is the linearized form of the LLG Eq.~(\ref{eqn1}),
\begin{eqnarray}
\partial_{t}\beta^{+}&=&\frac{\gamma'}{\mathcal{M}}\left(\textrm{i}-\alpha\right)\left[\left(D_{0}+D_{\textrm{nr}}\right)\beta^{+}+D_{\textrm{a}}\beta^{-}\right],\label{eqn5}
\\
\partial_{t}\beta^{-}&=&\frac{\gamma'}{\mathcal{M}}\left(-\textrm{i}-\alpha\right)\left[D^{\dag}_{\textrm{a}}\beta^{+}+\left(D_{0}-D_{\textrm{nr}}\right)\beta^{-}\right].\label{eqn6}
\end{eqnarray}

%
%

%

For details of the derivation see the Supplemental Material\cite{supp}. The term $D_{\textrm{nr}}$ in Eqs.~(\ref{eqn5})-(\ref{eqn6}) is responsible for the nonreciprocity of the SW spectrum\cite{Garst}. It accounts for the energy difference between magnons propagating in opposite directions in in-plane oriented ultrathin FM films\cite{Udvardi,Zakeri} with Dzyaloshinsky--Moriya interaction\cite{Dzyaloshinsky,Moriya} and the splitting between clockwise and counterclockwise modes of a single skyrmion\cite{Mruczkiewicz}.

Here we will focus on the effects of the anomalous term\cite{Schutte} $D_{\textrm{a}}$, which couples Eqs.~(\ref{eqn5})-(\ref{eqn6}) together. Equations (\ref{eqn5})-(\ref{eqn6}) may be rewritten as eigenvalue equations by assuming the time dependence
\begin{eqnarray}
\beta^{\pm}\left(\boldsymbol{r},t\right)=\textrm{e}^{-\textrm{i}\omega_{k}t}\beta^{\pm}_{k}\left(\boldsymbol{r}\right).\label{eqn10}
\end{eqnarray}

For $\alpha=0$, 
the spins will precess around their equilibrium direction $\tilde{\boldsymbol{S}}_{0}$. 
If the equations are uncoupled, the $\tilde{S}^{x}$ and $\tilde{S}^{y}$ variables describe circular polarization, similarly to the Larmor precession of a single spin in an external magnetic field. However, the spins are forced on an elliptic path due to the presence of the anomalous terms.

The effective damping parameter of mode $k$ is defined as
\begin{eqnarray}
\alpha_{k,\textrm{eff}}=\left|\frac{\textrm{Im}\:\omega_{k}}{\textrm{Re}\:\omega_{k}}\right|,\label{eqn11}
\end{eqnarray}
which is the inverse of the figure of merit introduced in Ref.~\cite{Tiwari}. Equation~(\ref{eqn11}) expresses the fact that $\textrm{Im}\:\omega_{k}$, the linewidth in resonance experiments or decay coefficient in time-resolved measurements, is proportional to the excitation frequency $\textrm{Re}\:\omega_{k}$.

%
%

\begin{figure}
\centering
\includegraphics[width=\columnwidth]{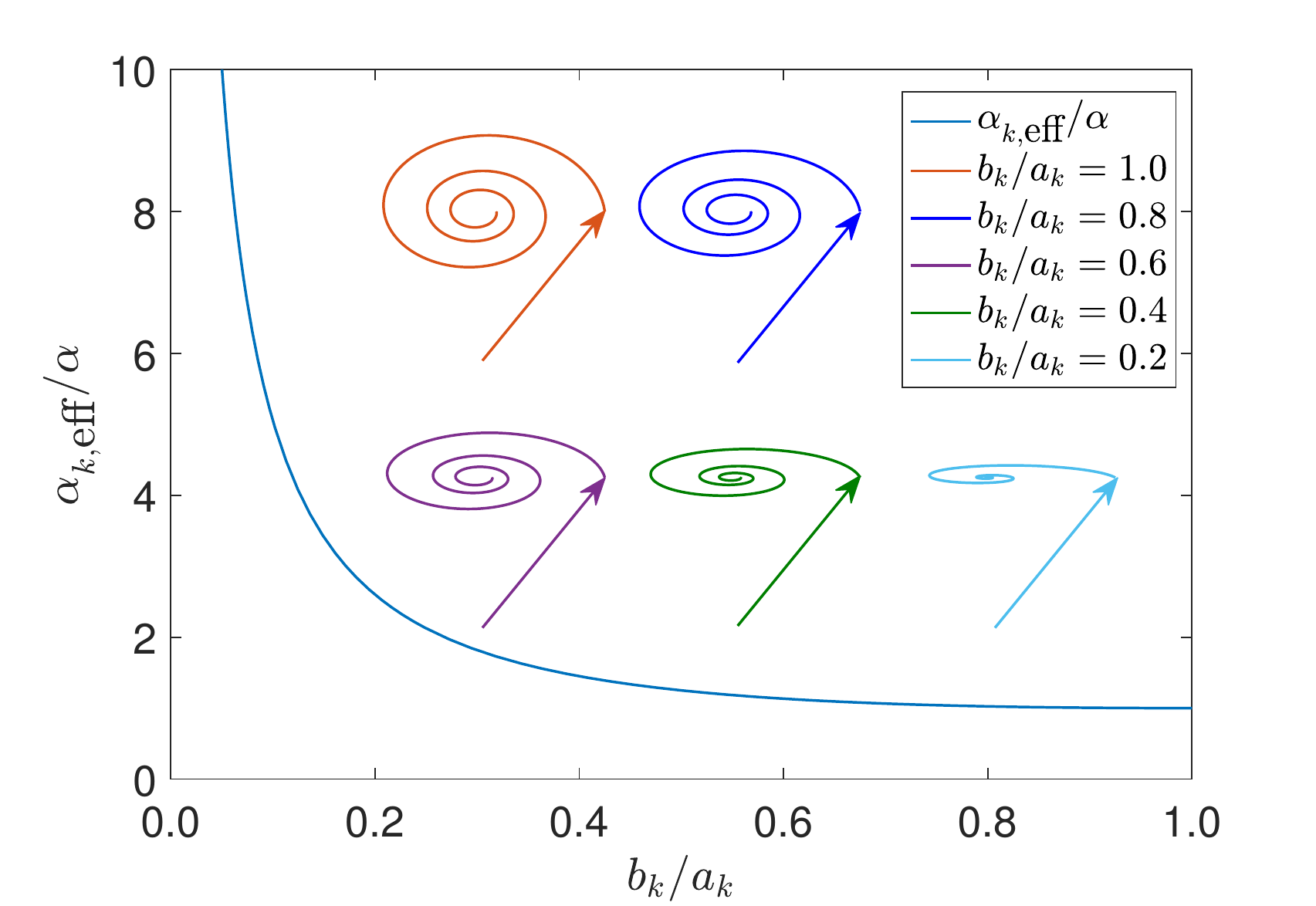}
\caption{Effective damping parameter $\alpha_{k,\textrm{eff}}$ as a function of inverse aspect ratio $b_{k}/a_{k}$ of the polarization ellipse, assuming constant $a_{k}$ and $b_{k}$ functions in Eq.~(\ref{eqn13}). Insets illustrate the precession for different values of $b_{k}/a_{k}$.\label{fig1}}
\end{figure}

Interestingly, there is a simple analytic expression connecting $\alpha_{k,\textrm{eff}}$ to the elliptic polarization of the modes at $\alpha=0$. For $\alpha\ll 1$, the effective damping may be expressed as
\begin{align}
\frac{\alpha_{k,\textrm{eff}}}{\alpha}\approx\frac{\bigintss\left|\beta_{k}^{-(0)}\left(\boldsymbol{r}\right)\right|^{2}+\left|\beta_{k}^{+(0)}\left(\boldsymbol{r}\right)\right|^{2}\textrm{d}\boldsymbol{r}}{\bigintss\left|\beta_{k}^{-(0)}\left(\boldsymbol{r}\right)\right|^{2}-\left|\beta_{k}^{+(0)}\left(\boldsymbol{r}\right)\right|^{2}\textrm{d}\boldsymbol{r}}=\frac{\bigintss a_{k}^{2}\left(\boldsymbol{r}\right)+b_{k}^{2}\left(\boldsymbol{r}\right)\textrm{d}\boldsymbol{r}}{\bigintss 2a_{k}\left(\boldsymbol{r}\right)b_{k}\left(\boldsymbol{r}\right)\textrm{d}\boldsymbol{r}},\label{eqn13}
\end{align}
where the $(0)$ superscript denotes that the eigenvectors $\beta^{\pm}_{k}\left(\boldsymbol{r}\right)$ defined in Eq.~(\ref{eqn10}) were calculated for $\alpha=0$, while $a_{k}\left(\boldsymbol{r}\right)$ and $b_{k}\left(\boldsymbol{r}\right)$ denote the semimajor and semiminor axes of the ellipse the spin variables $\tilde{S}^{x}\left(\boldsymbol{r}\right)$ and $\tilde{S}^{y}\left(\boldsymbol{r}\right)$ are precessing on in mode $k$. Details of the derivation are given in the Supplemental Material\cite{supp}. Note that an analogous expression for the uniform precession mode in FMs was derived in Ref.~\cite{Kambersky}. The main conclusion from Eq.~(\ref{eqn13}) is that $\alpha_{k,\textrm{eff}}$ will depend on the considered SW mode and it is always at least as high as the GD $\alpha$.
Although Eq.~(\ref{eqn13}) was obtained in the limit of low $\alpha$, numerical calculations indicate that the $\alpha_{k,\textrm{eff}}/\alpha$ ratio tends to increase for increasing values of $\alpha$; see the Supplemental Material\cite{supp} for an example. The enhancement of the damping from Eq.~(\ref{eqn13}) is shown in Fig.~\ref{fig1}, with the space-dependent $a_{k}\left(\boldsymbol{r}\right)$ and $b_{k}\left(\boldsymbol{r}\right)$ replaced by constants for simplicity. It can be seen that for more distorted polarization ellipses the spins get closer to the equilibrium direction after the same number of precessions, indicating a faster relaxation.

Since the appearance of the anomalous terms $D_{\textrm{a}}$ in Eqs.~(\ref{eqn5})-(\ref{eqn6}) forces the spins to precess on an elliptic path, it expresses that the system is not axially symmetric around the local spin directions in the equilibrium state denoted by $\boldsymbol{S}_{0}$. 
Such a symmetry breaking naturally occurs in any NC spin structure, implying a mode-dependent enhancement of the effective damping parameter in NC systems even within the phenomenological description of the LLG equation. Note that the NC structure also influences the electronic properties of the system, which can lead to a modification of the GD itself, see e.g. Ref.~\cite{Freimuth}.

%

In order to illustrate the enhanced and mode-dependent $\alpha_{k,\textrm{eff}}$, we calculate the magnons in isolated chiral skyrmions in a two-dimensional ultrathin film. The density of the Hamiltonian $\mathcal{H}$ reads\cite{Bogdanov2}
\begin{eqnarray}
h&=&\sum_{\alpha=x,y,z}\left[\mathcal{A}\left(\boldsymbol{\nabla}S^{\alpha}\right)^{2}\right]+\mathcal{K}\left(S^{z}\right)^{2}-\mathcal{M}BS^{z}\nonumber
\\
&&+\mathcal{D}\left(S^{z}\partial_{x}S^{x}-S^{x}\partial_{x}S^{z}+S^{z}\partial_{y}S^{y}-S^{y}\partial_{y}S^{z}\right),\label{eqn15}
\end{eqnarray}
with $\mathcal{A}$ the exchange stiffness, $\mathcal{D}$ the Dzyaloshinsky--Moriya interaction, $\mathcal{K}$ the anisotropy coefficient, and $B$ the external field.

In the following we will assume $\mathcal{D}>0$ and $B\ge 0$ without the loss of generality, see the Supplemental Material\cite{supp} for discussion. Using cylindrical coordinates $\left(r,\varphi\right)$ in real space and spherical coordinates $\boldsymbol{S}=\left(\sin\Theta\cos\Phi,\sin\Theta\sin\Phi,\cos\Theta\right)$ in spin space, the equilibrium profile of the isolated skyrmion will correspond to the cylindrically symmetric configuration $\Theta_{0}\left(r,\varphi\right)=\Theta_{0}\left(r\right)$ and $\Phi_{0}\left(r,\varphi\right)=\varphi$, the former satisfying
\begin{eqnarray}
&&\mathcal{A}\left(\partial_{r}^{2}\Theta_{0}+\frac{1}{r}\partial_{r}\Theta_{0}-\frac{1}{r^{2}}\sin\Theta_{0}\cos\Theta_{0}\right)+\mathcal{D}\frac{1}{r}\sin^{2}\Theta_{0}\nonumber
\\
&&+\mathcal{K}\sin\Theta_{0}\cos\Theta_{0}-\frac{1}{2}\mathcal{M}B\sin\Theta_{0}=0\label{eqn16}
\end{eqnarray}
with the boundary conditions $\Theta_{0}\left(0\right)=\pi,\Theta_{0}\left(\infty\right)=0$.

The operators in Eqs.~(\ref{eqn5})-(\ref{eqn6}) take the form (cf. Refs.~\cite{Schutte,Kravchuk,Lin2} and the Supplemental Material\cite{supp})
\begin{eqnarray}
D_{0}=&&-2\mathcal{A}\Bigg\{\boldsymbol{\nabla}^{2}+\frac{1}{2}\left[\left(\partial_{r}\Theta_{0}\right)^{2}-\frac{1}{r^{2}}\left(3\cos^{2}\Theta_{0}-1\right)\left(\partial_{\varphi}\Phi_{0}\right)^{2}\right]\Bigg\}\nonumber
\\
&&-\mathcal{D}\left(\partial_{r}\Theta_{0}+\frac{1}{r}3\sin\Theta_{0}\cos\Theta_{0}\partial_{\varphi}\Phi_{0}\right)\nonumber
\\
&&-\mathcal{K}\left(3\cos^{2}\Theta_{0}-1\right)+\mathcal{M}B\cos\Theta_{0},\label{eqn17}
\\
D_{\textrm{nr}}=&&\left(4\mathcal{A}\frac{1}{r^{2}}\cos\Theta_{0}\partial_{\varphi}\Phi_{0}-2\mathcal{D}\frac{1}{r}\sin\Theta_{0}\right)\left(-\textrm{i}\partial_{\varphi}\right),\label{eqn18}
\\
D_{\textrm{a}}=&&\mathcal{A}\left[\left(\partial_{r}\Theta_{0}\right)^{2}-\frac{1}{r^{2}}\sin^{2}\Theta_{0}\left(\partial_{\varphi}\Phi_{0}\right)^{2}\right]\nonumber
\\
&&+\mathcal{D}\left(\partial_{r}\Theta_{0}-\frac{1}{r}\sin\Theta_{0}\cos\Theta_{0}\partial_{\varphi}\Phi_{0}\right)+\mathcal{K}\sin^{2}\Theta_{0}.\label{eqn19}
\end{eqnarray}

Equation (\ref{eqn19}) demonstrates that the anomalous terms $D_{\textrm{a}}$ responsible for the enhancement of the effective damping can be attributed primarily to the NC arrangement ($\partial_{r}\Theta_{0}$ and $\partial_{\varphi}\Phi_{0}\equiv1$) and secondarily to the spins becoming canted with respect to the global out-of-plane symmetry axis ($\Theta_{0}\in\left\{0,\pi\right\}$) of the system. The $D_{\textrm{nr}}$ term introduces a nonreciprocity between modes with positive and negative values of the azimuthal quantum number $\left(-\textrm{i}\partial_{\varphi}\right)\rightarrow m$, preferring clockwise rotating modes ($m<0$) over counterclockwise rotating ones ($m>0$) following the sign convention of Refs.~\cite{Schutte,Mruczkiewicz}. Because $D_{0}$ 
and $D_{\textrm{nr}}$ 
depend on $m$ but $D_{\textrm{a}}$ does not, it is expected that the distortion of the SW polarization ellipse and consequently the effective damping will be more enhanced for smaller values of $\left|m\right|$.


\begin{figure}
\centering
\includegraphics[width=\columnwidth]{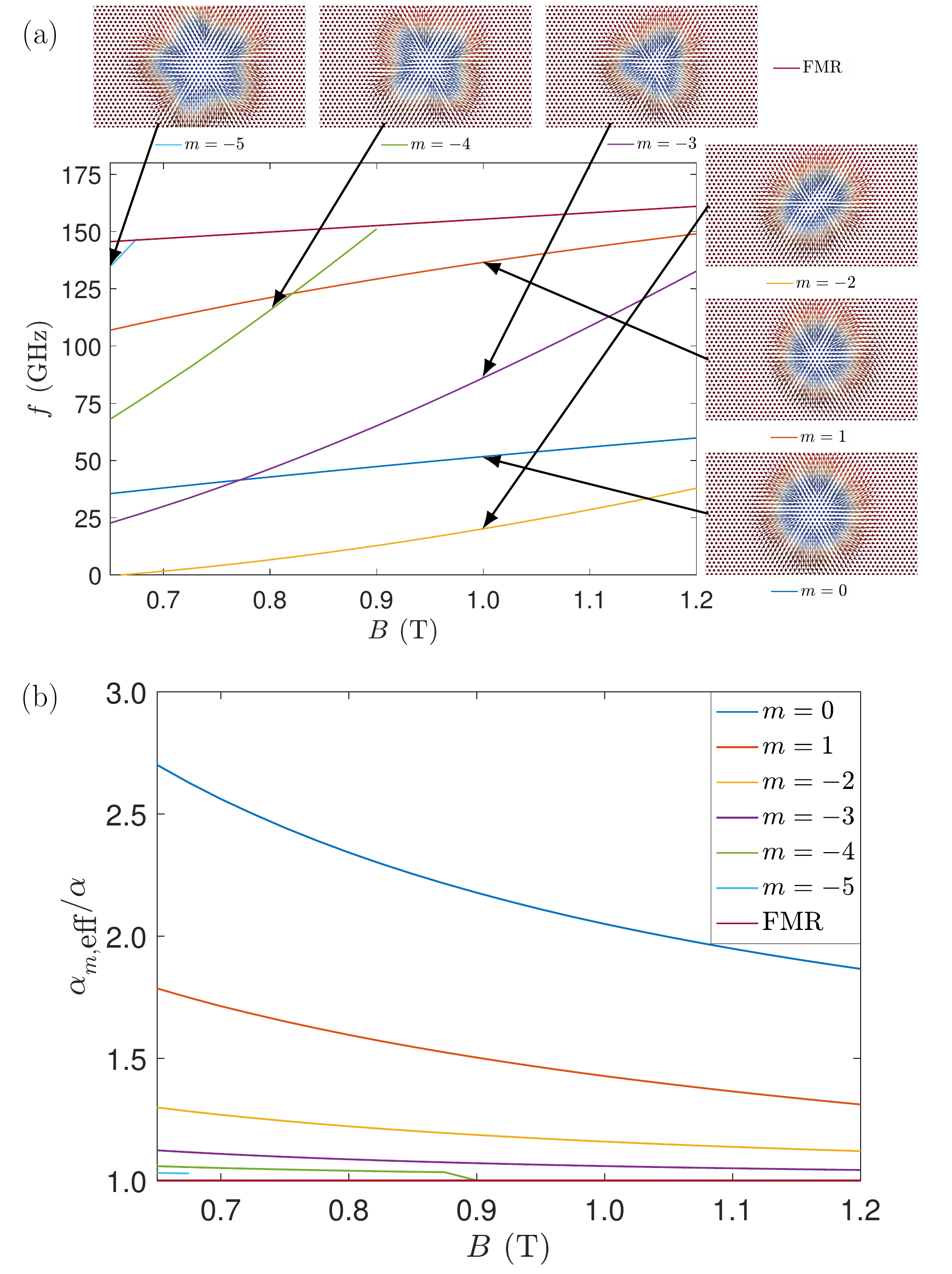}
\caption{Localized magnons in the isolated skyrmion, with the interaction parameters corresponding to the Pd/Fe/Ir(111) system\cite{Romming}: $\mathcal{A}=2.0\,\textrm{pJ/m},\mathcal{D}=-3.9\,\textrm{mJ/m}^{2},\mathcal{K}=-2.5\,\textrm{MJ/m}^{3},\mathcal{M}=1.1\,\textrm{MA/m}$. (a) Magnon frequencies $f=\omega/2\pi$ for $\alpha=0$. Illustrations display the shapes of the excitation modes visualized on the triangular lattice of Fe magnetic moments, with red and blue colors corresponding to positive and negative out-of-plane spin components, respectively. (b) Effective damping coefficients $\alpha_{m,\textrm{eff}}$, calculated from Eq.~(\ref{eqn13}).\label{fig2}}
\end{figure}

The different modes as a function of external field are shown in Fig.~\ref{fig2}(a), for the material parameters describing the Pd/Fe/Ir(111) system. The FMR mode at $\omega_{\textrm{FMR}}=\frac{\gamma}{\mathcal{M}}\left(\mathcal{M}B-2\mathcal{K}\right)$, describing a collective in-phase precession of the magnetization of the whole sample, separates the continuum and discrete parts of the spectrum, with the localized excitations of the isolated skyrmion located below the FMR frequency\cite{Schutte,Kravchuk}. We found a single localized mode for each $m\in\left\{0,1,-2,-3,-4,-5\right\}$ value, so in the following we will denote the excitation modes with the azimuthal quantum number. The $m=-1$ mode corresponds to the translation of the skyrmion on the field-polarized background, which is a zero-frequency Goldstone mode of the system and not shown in the figure. The $m=-2$ mode tends to zero around $B=0.65\,\textrm{T}$, indicating that isolated skyrmions become susceptible to elliptic deformations and subsequently cannot be stabilized at lower field values\cite{Bogdanov3}.


The values of $\alpha_{m,\textrm{eff}}$ calculated from Eq.~(\ref{eqn13}) 
for the different modes are summarized in Fig.~\ref{fig2}(b). It is important to note that for a skyrmion stabilized at a selected field value, the modes display widely different $\alpha_{m,\textrm{eff}}$ values, with the breathing mode $m=0$ being typically damped twice as strongly as the FMR mode. The effective damping tends to increase for lower field values, and decrease for increasing values of $\left|m\right|$, the latter property expected from the $m$-dependence of Eqs.~(\ref{eqn17})-(\ref{eqn19}) as discussed above. It is worth noting that the $\alpha_{m,\textrm{eff}}$ parameters are not directly related to the skyrmion size. We also performed the calculations for the parameters describing Ir|Co|Pt multilayers\cite{Moreau-Luchaire}, and for the significantly larger skyrmions in that system we obtained considerably smaller excitation frequencies, but quantitatively similar effective damping parameters; details are given in the Supplemental Material\cite{supp}.

\begin{figure}
\centering
\includegraphics[width=\columnwidth]{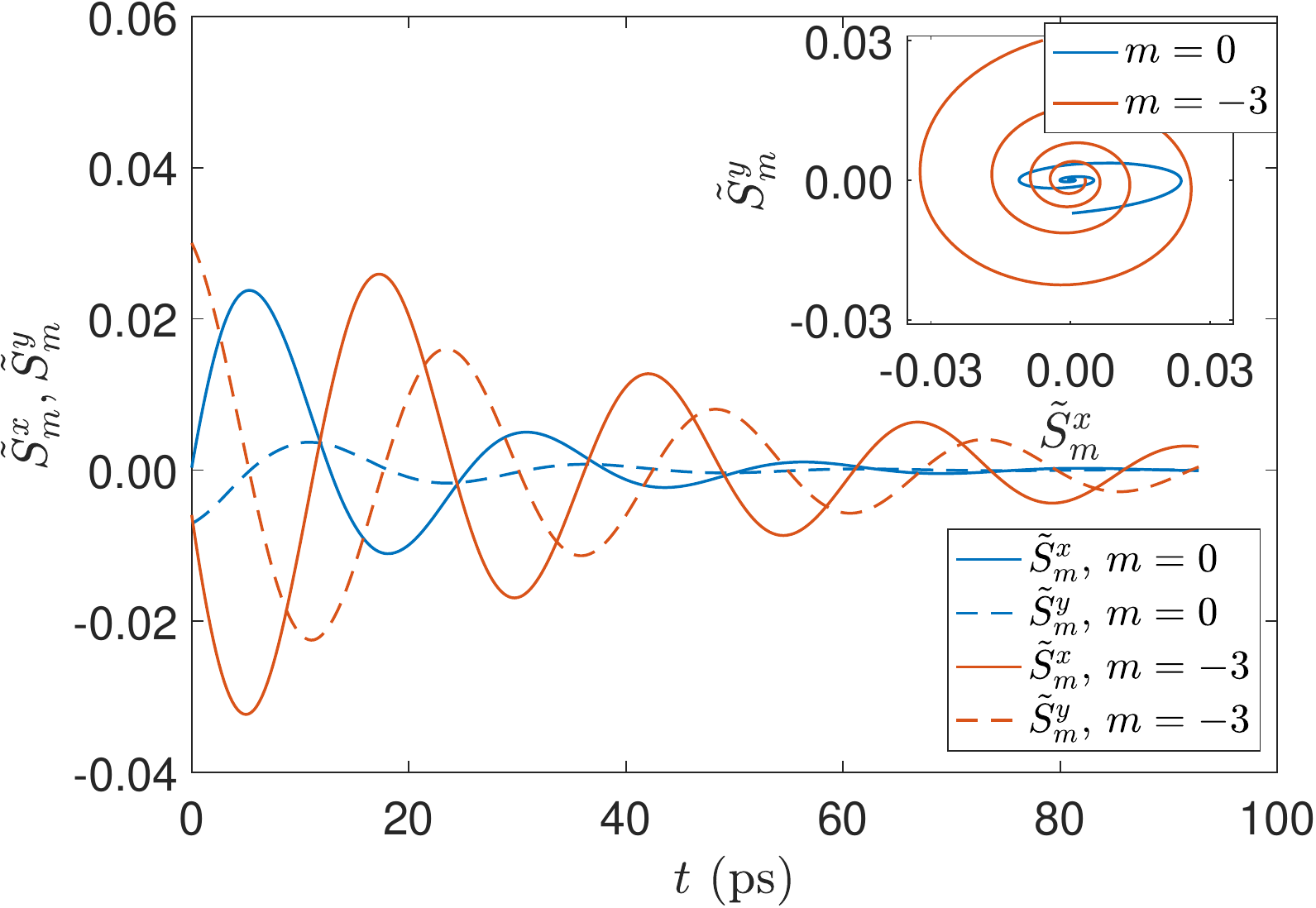}
\caption{Precession of a single spin in the skyrmion in the Pd/Fe/Ir(111) system in the $m=0$ and $m=-3$ modes at $B=0.75\,\textrm{T}$, from numerical simulations performed at $\alpha=0.1$. Inset shows the elliptic precession paths. From fitting the oscillations with Eq.~(\ref{eqn10}), we obtained $\left|\textrm{Re}\:\omega_{m=0}\right|/2\pi=39.22\,\textrm{GHz}$, $\left|\textrm{Im}\:\omega_{m=0}\right|=0.0608\,\textrm{ps}^{-1}$, $\alpha_{m=0,\textrm{eff}}=0.25$ and $\left|\textrm{Re}\:\omega_{m=-3}\right|/2\pi=40.31\,\textrm{GHz}$, $\left|\textrm{Im}\:\omega_{m=-3}\right|=0.0276\,\textrm{ps}^{-1}$, $\alpha_{m=-3,\textrm{eff}}=0.11$. \label{fig4}}
\end{figure}

The different effective damping parameters could possibly be determined experimentally by comparing the linewidths of the different excitation modes at a selected field value, or investigating the magnon decay over time. An example for the latter case is shown in Fig.~\ref{fig4}, displaying the precession of a single spin in the skyrmion, 
obtained from the numerical solution of the LLG Eq.~(\ref{eqn1}) with $\alpha=0.1$. At $B=0.75\,\textrm{T}$, the frequencies of the $m=0$ breathing and $m=-3$ triangular modes are close to each other (cf. Fig.~\ref{fig2}), but the former decays much faster. Because in the breathing mode the spin is following a significantly more distorted elliptic path (inset of Fig.~\ref{fig4}) than in the triangular mode, the different effective damping is also indicated by Eq.~(\ref{eqn13}).

In summary, it was demonstrated within the phenomenological description of the LLG equation that the effective damping parameter $\alpha_{\textrm{eff}}$ 
depends on the considered magnon mode. The $\alpha_{\textrm{eff}}$ assumes larger values if the
polarization ellipse is strongly distorted as expressed by Eq.~(\ref{eqn13}). Since NC magnetic structures provide an anisotropic environment for the spins, leading to a distortion of the precession path, they provide a natural choice for realizing different $\alpha_{\textrm{eff}}$ values within a single system. The results of the theory were demonstrated for isolated skyrmions with material parameters describing the Pd/Fe/Ir(111) system. 
The results presented here may stimulate further experimental or theoretical work on the effective damping in skyrmions, vortices, domain walls or spin spirals.

\begin{acknowledgments}

The authors would like to thank U. Atxitia and G. Meier for fruitful discussions. Financial support by the Alexander von Humboldt Foundation, by the Deutsche Forschungsgemeinschaft via SFB 668, by the European Union via the Horizon 2020 research and innovation program under Grant Agreement No. 665095 (MAGicSky), and by the National Research, Development and Innovation Office of Hungary under Project No. K115575 is gratefully acknowledged.

\end{acknowledgments}

\end{document}


\title{Supplemental Material to \\Effective damping enhancement in noncollinear spin structures}

\author{Levente R\'{o}zsa}
\email{rozsa.levente@physnet.uni-hamburg.de}
\author{Julian Hagemeister}
\author{Elena Y. Vedmedenko}
\author{Roland Wiesendanger}
\affiliation{Department of Physics, University of Hamburg, D-20355 Hamburg, Germany}
\date{\today}

\begin{abstract}

In the Supplemental Material the derivation of the linearized equations of motion and the effective damping parameter are discussed. Details of the numerical determination of the magnon modes in the continuum model and in atomistic spin dynamics simulations are also given.

\end{abstract}

\maketitle

\section{Linearized Landau--Lifshitz--Gilbert equation\label{secS1}}

Here we will derive the linearized form of the Landau--Lifshitz--Gilbert equation given in Eqs.~(2)-(3) of the main text and discuss the properties of the solutions. The calculation is similar to the undamped case, discussed in detail in e.g. Refs.~\cite{Schutte,Kravchuk,Lin2}. Given a spin configuration satisfying the equilibrium condition
\begin{eqnarray}
\boldsymbol{S}_{0}\times\boldsymbol{B}^{\textrm{eff}}=\boldsymbol{0},\label{eqnS1}
\end{eqnarray}
the local coordinate system with $\tilde{\boldsymbol{S}}_{0}=\left(0,0,1\right)$ may be introduced, and the Hamiltonian be expanded in the variables $\tilde{S}^{x}$ and $\tilde{S}^{y}$. The linear term must disappear because the expansion is carried out around an equilibrium state. The lowest-order nontrivial term is quadratic in the variables and will be designated as the spin wave Hamiltonian,
\begin{eqnarray}
\mathcal{H}_{\textrm{SW}}&=&\int h_{\textrm{SW}}\textrm{d}\boldsymbol{r},\label{eqnS2}
\\
h_{\textrm{SW}}&=&\frac{1}{2}\left[\begin{array}{cc} \tilde{S}^{x} & \tilde{S}^{y} \end{array}\right]\left[\begin{array}{cc} A_{1} & A_{2} \\ A_{2}^{\dag} & A_{3} \end{array}\right]\left[\begin{array}{c} \tilde{S}^{x} \\ \tilde{S}^{y} \end{array}\right]\nonumber
\\
&=&\frac{1}{2}\left(\tilde{\boldsymbol{S}}^{\perp}\right)^{T}H_{\textrm{SW}}\tilde{\boldsymbol{S}}^{\perp}.\label{eqnS3}
\end{eqnarray}

The operator $H_{\textrm{SW}}$ is self-adjoint for arbitrary equilibrium states. Here we will only consider cases where the equilibrium state is a local energy minimum, meaning that $H_{\textrm{SW}}\ge 0$; the magnon spectrum will only be well-defined in this case. Since $h_{\textrm{SW}}$ is obtained as an expansion of a real-valued energy density around the equilibrium state, and the spin variables are also real-valued, from the conjugate of Eq.~(\ref{eqnS3}) one gets $A_{1}=A_{1}^{*}$, $A_{2}=A_{2}^{*}$, and $A_{3}=A_{3}^{*}$.

The form of the Landau--Lifshitz--Gilbert Eq.~(1) in the main text may be rewritten in the local coordinates by simply replacing $\boldsymbol{S}$ by $\tilde{\boldsymbol{S}}_{0}$ everywhere, including the definition of the effective field $\boldsymbol{B}^{\textrm{eff}}$. The harmonic Hamiltonian $\mathcal{H}_{\textrm{SW}}$ in Eq.~(\ref{eqnS2}) leads to the linearized equation of motion
\begin{eqnarray}
\partial_{t}\tilde{\boldsymbol{S}}^{\perp}=\frac{\gamma'}{\mathcal{M}}\left(-\textrm{i}\sigma^{y}-\alpha\right)H_{\textrm{SW}}\tilde{\boldsymbol{S}}^{\perp},\label{eqnS4}
\end{eqnarray}
with $\sigma^{y}=\left[\begin{array}{cc} 0 & -\textrm{i} \\ \textrm{i} & 0 \end{array}\right]$ the Pauli matrix.

By replacing $\tilde{\boldsymbol{S}}^{\perp}\left(\boldsymbol{r},t\right)\rightarrow\tilde{\boldsymbol{S}}_{k}^{\perp}\left(\boldsymbol{r}\right)\textrm{e}^{-\textrm{i}\omega_{k}t}$ as usual, for $\alpha=0$ the eigenvalue equation
\begin{eqnarray}
\omega_{k}\tilde{\boldsymbol{S}}_{k}^{\perp}=\frac{\gamma}{\mathcal{M}}\sigma^{y}H_{\textrm{SW}}\tilde{\boldsymbol{S}}_{k}^{\perp}\label{eqnS5}
\end{eqnarray}
is obtained. If $H_{\textrm{SW}}$ has a strictly positive spectrum, then $H_{\textrm{SW}}^{-\frac{1}{2}}$ exists, and $\sigma^{y}H_{\textrm{SW}}$ has the same eigenvalues as $H_{\textrm{SW}}^{\frac{1}{2}}\sigma^{y}H_{\textrm{SW}}^{\frac{1}{2}}$. Since the latter is a self-adjoint matrix with respect to the standard scalar product on the Hilbert space, it has a real spectrum, consequently all $\omega_{k}$ eigenvalues are real. Note that the zero modes of $H_{\textrm{SW}}$, which commonly occur in the form of Goldstone modes due to the ground state breaking a continuous symmetry of the Hamiltonian, have to be treated separately. Finally, we mention that if the spin wave expansion is performed around an equilibrium state which is not a local energy minimum, the $\omega_{k}$ eigenvalues may become imaginary, meaning that the linearized Landau--Lifshitz--Gilbert equation will describe a divergence from the unstable equilibrium state instead of a precession around it.

Equations (2)-(3) in the main text may be obtained by introducing the variables $\beta^{\pm}=\tilde{S}^{x}\pm\textrm{i}\tilde{S}^{y}$ as described there. The connection between $H_{\textrm{SW}}$ and the operators $D_{0}$, $D_{\textrm{nr}}$, and $D_{\textrm{a}}$ is given by
\begin{eqnarray}
D_{0}&=&\frac{1}{2}\left(A_{1}+A_{3}\right),\label{eqnS6}
\\
D_{\textrm{nr}}&=&\frac{1}{2}\textrm{i}\left(A_{2}^{\dag}-A_{2}\right),\label{eqnS7}
\\
D_{\textrm{a}}&=&\frac{1}{2}\left[A_{1}-A_{3}+\textrm{i}\left(A_{2}^{\dag}+A_{2}\right)\right].\label{eqnS8}
\end{eqnarray}

An important symmetry property of Eqs. (2)-(3) in the main text is that if $\left(\beta^{+},\beta^{-}\right)=\left(\beta^{+}_{k}\textrm{e}^{-\textrm{i}\omega_{k}t},\beta^{-}_{k}\textrm{e}^{-\textrm{i}\omega_{k}t}\right)$ is an eigenmode of the equations, then $\left(\beta^{+},\beta^{-}\right)=\left(\left(\beta^{-}_{k}\right)^{*}\textrm{e}^{\textrm{i}\omega^{*}_{k}t},\left(\beta^{+}_{k}\right)^{*}\textrm{e}^{\textrm{i}\omega^{*}_{k}t}\right)$ is another solution. Following Refs.~\cite{Schutte,Lin2}, this can be attributed to the particle-hole symmetry of the Hamiltonian, which also holds in the presence of the damping term. From these two solutions mentioned above, the real-valued time evolution of the variables $\tilde{S}^{x},\tilde{S}^{y}$ may be expressed as
\begin{eqnarray}
\tilde{S}_{k}^{x}&=&\textrm{e}^{\textrm{Im}\:\omega_{k}t}\cos\left(\varphi_{+,k}-\textrm{Re}\:\omega_{k}t\right)\left|\beta_{k}^{+}+\beta_{k}^{-}\right|,\label{eqnS9}
\\
\tilde{S}_{k}^{y}&=&\textrm{e}^{\textrm{Im}\:\omega_{k}t}\sin\left(\varphi_{-,k}-\textrm{Re}\:\omega_{k}t\right)\left|\beta_{k}^{+}-\beta_{k}^{-}\right|,\label{eqnS10}
\end{eqnarray}
with $\varphi_{\pm,k}=\arg\left(\beta_{k}^{+}\pm\beta_{k}^{-}\right)$. As mentioned above, the $\textrm{Im}\:\omega_{k}$ terms are zero in the absence of damping close to a local energy minimum, and $\textrm{Im}\:\omega_{k}<0$ is implied by the fact that the Landau--Lifshitz--Gilbert equation describes energy dissipation, which in the linearized case corresponds to relaxation towards the local energy minimum. In the absence of damping, the spins will precess on an ellipse defined by the equation
\begin{eqnarray}
&&\frac{\left(\tilde{S}_{k}^{x}\right)^{2}}{\left|\beta_{k}^{+(0)}+\beta_{k}^{-(0)}\right|^{2}\cos^{2}\left(\varphi_{+,k}-\varphi_{-,k}\right)}\nonumber
\\
&&+\frac{2\tilde{S}_{k}^{x}\tilde{S}_{k}^{y}\sin\left(\varphi_{+,k}-\varphi_{-,k}\right)}{\left|\beta_{k}^{+(0)}-\beta_{k}^{-(0)}\right|\left|\beta_{k}^{+(0)}+\beta_{k}^{-(0)}\right|\cos^{2}\left(\varphi_{+,k}-\varphi_{-,k}\right)}\nonumber
\\
&&+\frac{\left(\tilde{S}_{k}^{y}\right)^{2}}{\left|\beta_{k}^{+(0)}-\beta_{k}^{-(0)}\right|^{2}\cos^{2}\left(\varphi_{+,k}-\varphi_{-,k}\right)}=1,\label{eqnS11}
\end{eqnarray}
where the superscript $(0)$ indicates $\alpha=0$. The semimajor and semiminor axes of the ellipse $a_{k}$ and $b_{k}$ may be expressed from Eq.~(\ref{eqnS11}) as
\begin{eqnarray}
a_{k}b_{k}&=&\left|\left|\beta_{k}^{-(0)}\right|^{2}-\left|\beta_{k}^{+(0)}\right|^{2}\right|,\label{eqnS12}
\\
a^{2}_{k}+b^{2}_{k}&=&2\left(\left|\beta_{k}^{-(0)}\right|^{2}+\left|\beta_{k}^{+(0)}\right|^{2}\right).\label{eqnS13}
\end{eqnarray}

Note that $\beta_{k}^{+}$ and $\beta_{k}^{-}$, consequently the parameters of the precessional ellipse $a_{k}$ and $b_{k}$, are functions of the spatial position $\boldsymbol{r}$.

\section{Calculation of the effective damping parameter from perturbation theory\label{secS2}}

Here we derive the expression for the effective damping parameter $\alpha_{\textrm{eff}}$ given in Eq.~(6) of the main text. By introducing $\boldsymbol{\beta}_{k}=\left(\beta_{k}^{+},-\beta_{k}^{-}\right)$,
\begin{eqnarray}
\boldsymbol{D}=\left[\begin{array}{cc} D_{0}+D_{\textrm{nr}} & -D_{\textrm{a}} \\ -D^{\dag}_{\textrm{a}} & D_{0}-D_{\textrm{nr}} \end{array}\right],\label{eqnS14}
\end{eqnarray}
and using the Pauli matrix $\sigma^{z}=\left[\begin{array}{cc} 1 & 0 \\ 0 & -1 \end{array}\right]$, Eqs.~(2)-(3) in the main text may be rewritten as
\begin{eqnarray}
-\omega_{k}\sigma^{z}\boldsymbol{\beta}_{k}=\frac{\gamma'}{\mathcal{M}}\left(\boldsymbol{D}+\textrm{i}\alpha\sigma^{z}\boldsymbol{D}\right)\boldsymbol{\beta}_{k}\label{eqnS15}
\end{eqnarray}
in the frequency domain. Following standard perturbation theory, we expand the eigenvalues $\omega_{k}$ and the eigenvectors $\boldsymbol{\beta}_{k}$ in the parameter $\alpha\ll 1$. For the zeroth-order terms one gets
\begin{eqnarray}
-\omega^{(0)}_{k}\sigma^{z}\boldsymbol{\beta}^{(0)}_{k}=\frac{\gamma}{\mathcal{M}}\boldsymbol{D}\boldsymbol{\beta}^{(0)}_{k},\label{eqnS16}
\end{eqnarray}
with real $\omega^{(0)}_{k}$ eigenvalues as discussed in Sec.~\ref{secS1}. The first-order terms read
\begin{align}
&-\omega^{(0)}_{k}\left<\boldsymbol{\beta}^{(0)}_{k}\right|\sigma^{z}\left|\boldsymbol{\beta}^{(1)}_{k}\right>-\omega^{(1)}_{k}\left<\boldsymbol{\beta}^{(0)}_{k}\right|\sigma^{z}\left|\boldsymbol{\beta}^{(0)}_{k}\right>\nonumber
\\
&=\frac{\gamma}{\mathcal{M}}\left<\boldsymbol{\beta}^{(0)}_{k}\right|\boldsymbol{D}\left|\boldsymbol{\beta}^{(1)}_{k}\right>+\textrm{i}\alpha\frac{\gamma}{\mathcal{M}}\left<\boldsymbol{\beta}^{(0)}_{k}\right|\sigma^{z}\boldsymbol{D}\left|\boldsymbol{\beta}^{(0)}_{k}\right>,\label{eqnS17}
\end{align}
after taking the scalar product with $\boldsymbol{\beta}^{(0)}_{k}$. The first terms on both sides cancel by letting $\boldsymbol{D}$ act to the left, then using Eq.~(\ref{eqnS16}) and the fact that the $\omega^{(0)}_{k}$ are real. By applying Eq.~(\ref{eqnS16}) to the remaining term on the right-hand side one obtains
\begin{eqnarray}
\omega^{(1)}_{k}=-\textrm{i}\alpha\omega^{(0)}_{k}\frac{\bigintss\left|\beta_{k}^{-(0)}\right|^{2}+\left|\beta_{k}^{+(0)}\right|^{2}\textrm{d}\boldsymbol{r}}{\bigintss\left|\beta_{k}^{-(0)}\right|^{2}-\left|\beta_{k}^{+(0)}\right|^{2}\textrm{d}\boldsymbol{r}},\label{eqnS18}
\end{eqnarray}
by writing in the definition of the scalar product. By using the definition $\alpha_{k,\textrm{eff}}=\left|\textrm{Im}\omega_{k}/\textrm{Re}\omega_{k}\right|\approx\left|\omega^{(1)}_{k}/\omega^{(0)}_{k}\right|$ and substituting Eqs.~(\ref{eqnS12})-(\ref{eqnS13}) into Eq.~(\ref{eqnS18}), one arrives at Eq.~(6) in the main text as long as $\left|\beta_{k}^{-(0)}\right|^{2}-\left|\beta_{k}^{+(0)}\right|^{2}$ does not change sign under the integral.

\begin{figure}
\centering
\includegraphics[width=\columnwidth]{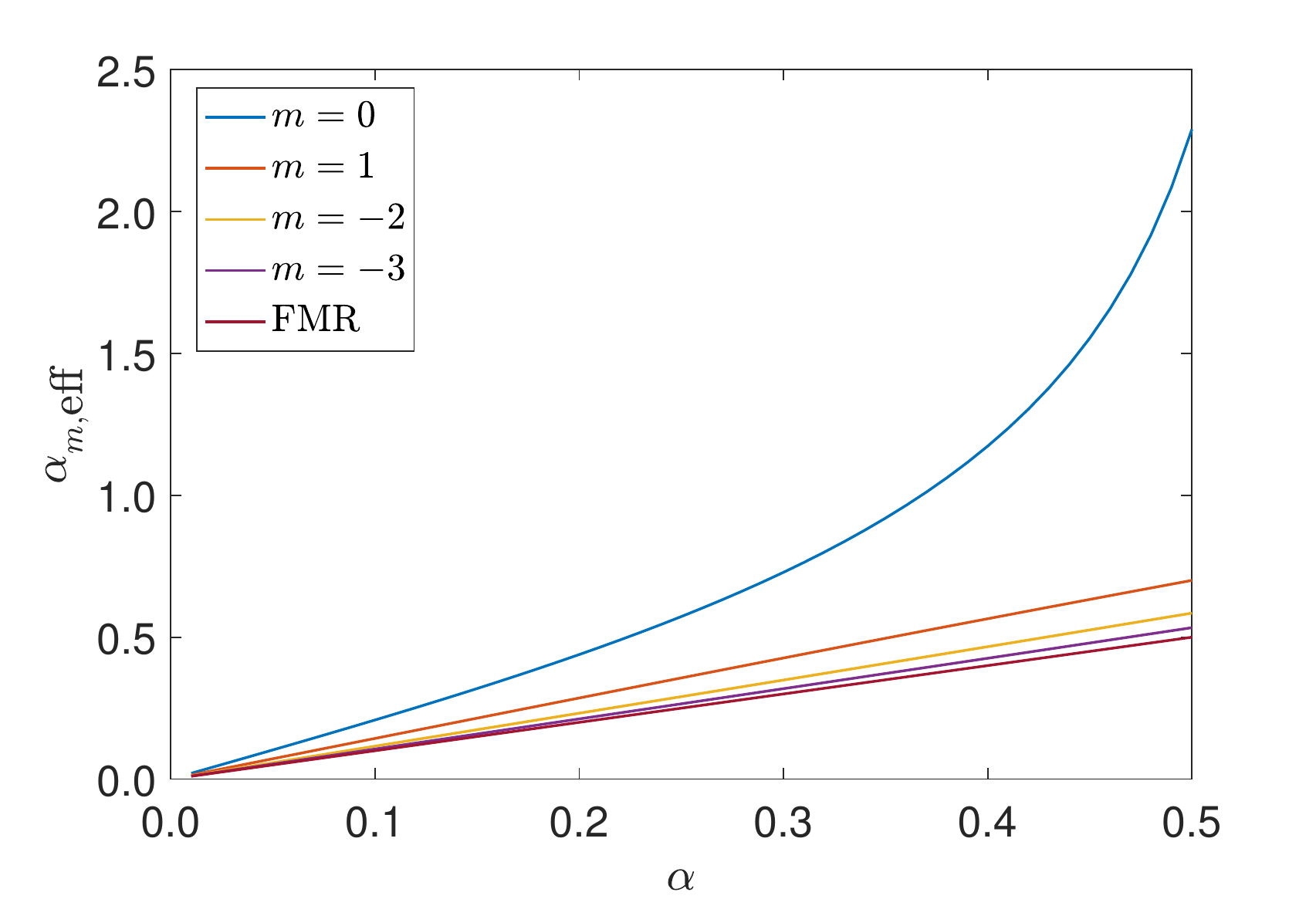}
\caption{Effective damping coefficients $\alpha_{m,\textrm{eff}}$ of the isolated skyrmion in the Pd/Fe/Ir(111) system at $B=1\,\textrm{T}$, calculated from the numerical solution of the linearized Landau--Lifshitz--Gilbert equation (\ref{eqnS15}), as a function of the Gilbert damping parameter $\alpha$.\label{figS1}}
\end{figure}

It is worthwhile to investigate for which values of $\alpha$ does first-order perturbation theory give a good estimate for $\alpha_{k,\textrm{eff}}$ calculated from the exact solution of the linearized equations of motion, Eq.~(\ref{eqnS15}). In the materials where the excitations of isolated skyrmions or skyrmion lattices were investigated, significantly different values of $\alpha$ have been found. For example, intrinsic Gilbert damping parameters of $\alpha=0.02$-$0.04$ were determined experimentally for bulk chiral magnets MnSi and Cu$_{2}$OSeO$_{3}$\cite{Schwarze}, $\alpha=0.28$ was deduced for FeGe\cite{Beg}, and a total damping of $\alpha_{\textrm{tot}}=0.105$ was obtained for Ir/Fe/Co/Pt magnetic multilayers\cite{Satywali}, where the latter value also includes various effects beyond the Landau--Lifshitz--Gilbert description. Figure~\ref{figS1} displays the dependence of $\alpha_{m,\textrm{eff}}$ on $\alpha$ for the eigenmodes of the isolated skyrmion in the Pd/Fe/Ir(111) system, shown in Fig.~2 of the main text. Most of the modes show a linear correspondence between the two quantities with different slopes in the displayed parameter range, in agreement with Eq.~(6) in the main text. For the breathing mode $m=0$ the convex shape of the curve indicates that the effective damping parameter becomes relatively even larger than the perturbative expression Eq.~(6) as $\alpha$ is increased.

\section{Eigenmodes of the isolated skyrmion\label{secS3}}

Here we discuss the derivation of the skyrmion profile Eq.~(8) and the operators in Eqs.~(9)-(11) of the main text. The energy density Eq.~(7) in polar coordinates reads
\begin{align}
&h=\mathcal{A}\bigg[\left(\partial_{r}\Theta\right)^{2}+\sin^{2}\Theta\left(\partial_{r}\Phi\right)^{2}+\frac{1}{r^{2}}\left(\partial_{\varphi}\Theta\right)^{2}\nonumber
\\
&+\frac{1}{r^{2}}\sin^{2}\Theta\left(\partial_{\varphi}\Phi\right)^{2}\bigg]+\mathcal{D}\bigg[\cos\left(\varphi-\Phi\right)\partial_{r}\Theta\nonumber
\\
&-\frac{1}{r}\sin\left(\varphi-\Phi\right)\partial_{\varphi}\Theta+\sin\Theta\cos\Theta\sin\left(\varphi-\Phi\right)\partial_{r}\Phi\nonumber
\\
&+\frac{1}{r}\sin\Theta\cos\Theta\cos\left(\varphi-\Phi\right)\partial_{\varphi}\Phi\bigg]+\mathcal{K}\cos^{2}\Theta-\mathcal{M}B\cos\Theta.\label{eqnS19}
\end{align}

The Landau--Lifshitz--Gilbert Eq.~(1) may be rewritten as
\begin{eqnarray}
\sin\Theta\partial_{t}\Theta&=&\gamma'B^{\Phi}+\alpha\gamma'\sin\Theta B^{\Theta},\label{eqnS20}
\\
\sin\Theta\partial_{t}\Phi&=&-\gamma'B^{\Theta}+\alpha\gamma'\frac{1}{\sin\Theta}B^{\Phi},\label{eqnS21}
\end{eqnarray}
with
\begin{align}
B^{\chi}=&-\frac{1}{\mathcal{M}}\frac{\delta \mathcal{H}}{\delta \chi}\nonumber
\\
=&-\frac{1}{\mathcal{M}}\left[-\frac{1}{r}\partial_{r}\left(r\frac{\partial h}{\partial \left(\partial_{r}\chi\right)}\right)-\partial_{\varphi}\frac{\partial h}{\partial \left(\partial_{\varphi}\chi\right)}+\frac{\partial h}{\partial\chi}\right],\label{eqnS22}
\end{align}
where $\chi$ stands for $\Theta$ or $\Phi$. Note that in this form it is common to redefine $B^{\Phi}$ to include the $1/\sin\Theta$ factor in Eq.~(\ref{eqnS21})\cite{Roma}. The first variations of $\mathcal{H}$ from Eq.~(\ref{eqnS19}) may be expressed as
\begin{align}
\frac{\delta \mathcal{H}}{\delta \Theta}=&-2\mathcal{A}\left\{\boldsymbol{\nabla}^{2}\Theta-\sin\Theta\cos\Theta\left[\left(\partial_{r}\Phi\right)^{2}+\frac{1}{r^{2}}\left(\partial_{\varphi}\Phi\right)^{2}\right]\right\}\nonumber
\\
&-2\mathcal{K}\sin\Theta\cos\Theta+\mathcal{M}B\sin\Theta\nonumber
\\
&-2\mathcal{D}\sin^{2}\Theta\left[\sin\left(\varphi-\Phi\right)\partial_{r}\Phi+\cos\left(\varphi-\Phi\right)\frac{1}{r}\partial_{\varphi}\Phi\right],\label{eqnS23}
\\
\frac{\delta \mathcal{H}}{\delta \Phi}=&-2\mathcal{A}\left\{\sin^{2}\Theta\boldsymbol{\nabla}^{2}\Phi+\sin2\Theta\left[\partial_{r}\Theta\partial_{r}\Phi+\frac{1}{r^{2}}\partial_{\varphi}\Theta\partial_{\varphi}\Phi\right]\right\}\nonumber
\\
&+2\mathcal{D}\sin^{2}\Theta\left[\sin\left(\varphi-\Phi\right)\partial_{r}\Theta+\cos\left(\varphi-\Phi\right)\frac{1}{r}\partial_{\varphi}\Theta\right],\label{eqnS24}
\end{align}

The equilibrium condition Eq.~(8) in the main text may be obtained by setting $\partial_{t}\Theta=\partial_{t}\Phi=0$ in Eqs.~(\ref{eqnS20})-(\ref{eqnS21}) and assuming cylindrical symmetry, $\Theta_{0}\left(r,\varphi\right)=\Theta_{0}\left(r\right)$ and $\Phi_{0}\left(r,\varphi\right)=\varphi$. In the main text $\mathcal{D}>0$ and $B\ge 0$ were assumed. Choosing $\mathcal{D}<0$ switches the helicity of the structure to $\Phi_{0}=\varphi+\pi$, in which case $\mathcal{D}$ should be replaced by $\left|\mathcal{D}\right|$ in Eq.~(8). For the background magnetization pointing in the opposite direction $B\le 0$, one obtains the time-reversed solutions with $\Theta_{0}\rightarrow\pi-\Theta_{0},\Phi_{0}\rightarrow\Phi_{0}+\pi,B\rightarrow-B$. Time reversal also reverses clockwise and counterclockwise rotating eigenmodes; however, the above transformations do not influence the magnitudes of the excitation frequencies. Finally, we note that the frequencies remain unchanged even if the form of the Dzyaloshinsky--Moriya interaction in Eq.~(\ref{eqnS19}), describing N\'{e}el-type skyrmions common in ultrathin films and multilayers, is replaced by an expression that prefers Bloch-type skyrmions occurring in bulk helimagnets -- see Ref.~\cite{Lin2} for details.

For determining the linearized equations of motion, one can proceed by switching to the local coordinate system as discussed in Sec.~\ref{secS1} and Refs.~\cite{Schutte,Lin2}. Alternatively, they can also directly be derived from Eqs.~(\ref{eqnS20})-(\ref{eqnS21}) by introducing $\Theta=\Theta_{0}+\tilde{S}^{x}, \Phi=\Phi_{0}+\frac{1}{\sin\Theta_{0}}\tilde{S}^{y}$ and expanding around the skyrmion profile from Eq.~(8) up to first order in $\tilde{S}^{x},\tilde{S}^{y}$ -- see also Ref.~\cite{Kravchuk}. The operators in Eq.~(\ref{eqnS3}) read
\begin{align}
A_{1}=&-2\mathcal{A}\left(\boldsymbol{\nabla}^{2}-\frac{1}{r^{2}}\cos2\Theta_{0}\left(\partial_{\varphi}\Phi_{0}\right)^{2}\right)\nonumber
\\
&-2\mathcal{D}\frac{1}{r}\sin2\Theta_{0}\partial_{\varphi}\Phi_{0}-2\mathcal{K}\cos2\Theta_{0}+\mathcal{M}B\cos\Theta_{0},\label{eqnS25}
\\
A_{2}=&4\mathcal{A}\frac{1}{r^{2}}\cos\Theta_{0}\partial_{\varphi}\Phi_{0}\partial_{\varphi}-2\mathcal{D}\frac{1}{r}\sin\Theta_{0}\partial_{\varphi},\label{eqnS26}
\\
A_{3}=&-2\mathcal{A}\left\{\boldsymbol{\nabla}^{2}+\left[\left(\partial_{r}\Theta_{0}\right)^{2}-\frac{1}{r^{2}}\cos^{2}\Theta_{0}\left(\partial_{\varphi}\Phi_{0}\right)^{2}\right]\right\}\nonumber
\\
&-2\mathcal{D}\left(\partial_{r}\Theta_{0}+\frac{1}{r}\sin\Theta_{0}\cos\Theta_{0}\partial_{\varphi}\Phi_{0}\right)\nonumber
\\
&-2\mathcal{K}\cos^{2}\Theta_{0}+\mathcal{M}B\cos\Theta_{0},\label{eqnS27}
\end{align}
which leads directly to Eqs.~(9)-(11) in the main text via Eqs.~(\ref{eqnS6})-(\ref{eqnS8}).

The excitation frequencies of the ferromagnetic state may be determined by setting $\Theta_{0}\equiv 0$ in Eqs.~(9)-(11) in the main text. In this case, the eigenvalues and eigenvectors can be calculated analytically\cite{Schutte},
\begin{eqnarray}
&&\omega_{k,m}=\frac{\gamma'}{\mathcal{M}}\left(1-\textrm{i}\alpha\right)\left[2\mathcal{A}k^{2}-2\mathcal{K}+\mathcal{M}B\right],\label{eqnS27a}
\\
&&\left(\beta^{+}_{k,m}\left(r\right),\beta^{-}_{k,m}\left(r\right)\right)=\left(0,J_{m-1}\left(kr\right)\right),\label{eqnS27b}
\end{eqnarray}
with $J_{m-1}$ the Bessel function of the first kind, appearing due to the solutions being regular at the origin. Equation~(\ref{eqnS27a}) demonstrates that the lowest-frequency excitation of the background is the ferromagnetic resonance frequency $\omega_{\textrm{FMR}}=\frac{\gamma}{\mathcal{M}}\left(\mathcal{M}B-2\mathcal{K}\right)$ at $\alpha=0$. Since the anomalous term $D_{\textrm{a}}$ disappears in the out-of-plane magnetized ferromagnetic state, all spin waves will be circularly polarized, see Eq.~(\ref{eqnS27b}), and the effective damping parameter will always coincide with the Gilbert damping.

Regarding the excitations of the isolated skyrmion, for $\alpha=0$ the linearized equations of motion in Eq. (\ref{eqnS15}) are real-valued; consequently, $\beta^{\pm}_{k,m}\left(r\right)$ can be chosen to be real-valued. In this case Eqs.~(\ref{eqnS9})-(\ref{eqnS10}) take the form
\begin{eqnarray}
\tilde{S}_{k,m}^{x}&=&\cos\left(m\varphi-\omega_{k,m}t\right)\left(\beta^{+}_{k,m}\left(r\right)+\beta^{-}_{k,m}\left(r\right)\right)\!,\:\:\:\:\:\:\label{eqnS27c}
\\
\tilde{S}_{k,m}^{y}&=&\sin\left(m\varphi-\omega_{k,m}t\right)\left(\beta^{+}_{k,m}\left(r\right)-\beta^{-}_{k,m}\left(r\right)\right)\!.\:\:\:\:\:\:\label{eqnS27d}
\end{eqnarray}

This means that modes with $\omega_{k,m}>0$ for $m>0$ will rotate counterclockwise, that is, the contours with constant $\tilde{S}_{k,m}^{x}$ and $\tilde{S}_{k,m}^{y}$ will move towards higher values of $\varphi$ as $t$ is increased, while the modes with $\omega_{k,m}>0$ for $m<0$ will rotate clockwise. Modes with $m=0$ correspond to breathing excitations. This sign convention for $m$ was used when designating the localized modes of the isolated skyrmion in the main text, and the $k$ index was dropped since only a single mode could be observed below the ferromagnetic resonance frequency for each value of $m$.

\section{Numerical solution of the eigenvalue equations\label{secS4}}

The linearized Landau--Lifshitz--Gilbert equation for the isolated skyrmion, Eqs.~(2)-(3) with the operators Eqs.~(9)-(11) in the main text, were solved numerically by a finite-difference method. First the equilibrium profile was determined from Eq.~(8) using the shooting method for an initial approximation, then obtaining the solution on a finer grid via finite differences. For the calculations we used dimensionless parameters (cf. Ref.~\cite{Leonov}),
\begin{eqnarray}
\mathcal{A}_{\textrm{dl}}&=&1,\label{eqnS28}
\\
\mathcal{D}_{\textrm{dl}}&=&1,\label{eqnS29}
\\
\mathcal{K}_{\textrm{dl}}&=&\frac{\mathcal{K}\mathcal{A}}{\mathcal{D}^{2}},\label{eqnS30}
\\
\left(\mathcal{M}B\right)_{\textrm{dl}}&=&\frac{\mathcal{M}B\mathcal{A}}{\mathcal{D}^{2}},\label{eqnS31}
\\
r_{\textrm{dl}}&=&\frac{\left|\mathcal{D}\right|}{\mathcal{A}}r,\label{eqnS32}
\\
\omega_{\textrm{dl}}&=&\frac{\mathcal{M}\mathcal{A}}{\gamma\mathcal{D}^{2}}\omega.\label{eqnS33}
\end{eqnarray}

The equations were solved in a finite interval for $r_{\textrm{dl}}\in\left[0,R\right]$, with the boundary conditions $\Theta_{0}\left(0\right)=\pi,\Theta_{0}\left(R\right)=0$. For the results presented in Fig.~2 in the main text the value of $R=30$ was used. It was confirmed by modifying $R$ that the skyrmion shape and the frequencies of the localized modes were not significantly affected by the boundary conditions. However, the frequencies of the modes above the ferromagnetic resonance frequency $\omega_{\textrm{FMR}}=\frac{\gamma}{\mathcal{M}}\left(\mathcal{M}B-2\mathcal{K}\right)$ did change as a function of $R$, since these modes are extended over the ferromagnetic background -- see Eqs.~(\ref{eqnS27a})-(\ref{eqnS27b}). Furthermore, in the infinitely extended system the equations of motion include a Goldstone mode with $\left(\beta^{+}_{m=-1},\beta^{-}_{m=-1}\right)=\left(-\frac{1}{r}\sin\Theta_{0}-\partial_{r}\Theta_{0},\frac{1}{r}\sin\Theta_{0}-\partial_{r}\Theta_{0}\right)$, corresponding to the translation of the skyrmion on the collinear background\cite{Schutte}. This mode obtains a finite frequency in the numerical calculations due to the finite value of $R$ and describes a slow clockwise gyration of the skyrmion. However, this frequency is not shown in Fig.~3 of the main text because it is only created by boundary effects.

%

\begin{figure}
\centering
\includegraphics[width=\columnwidth]{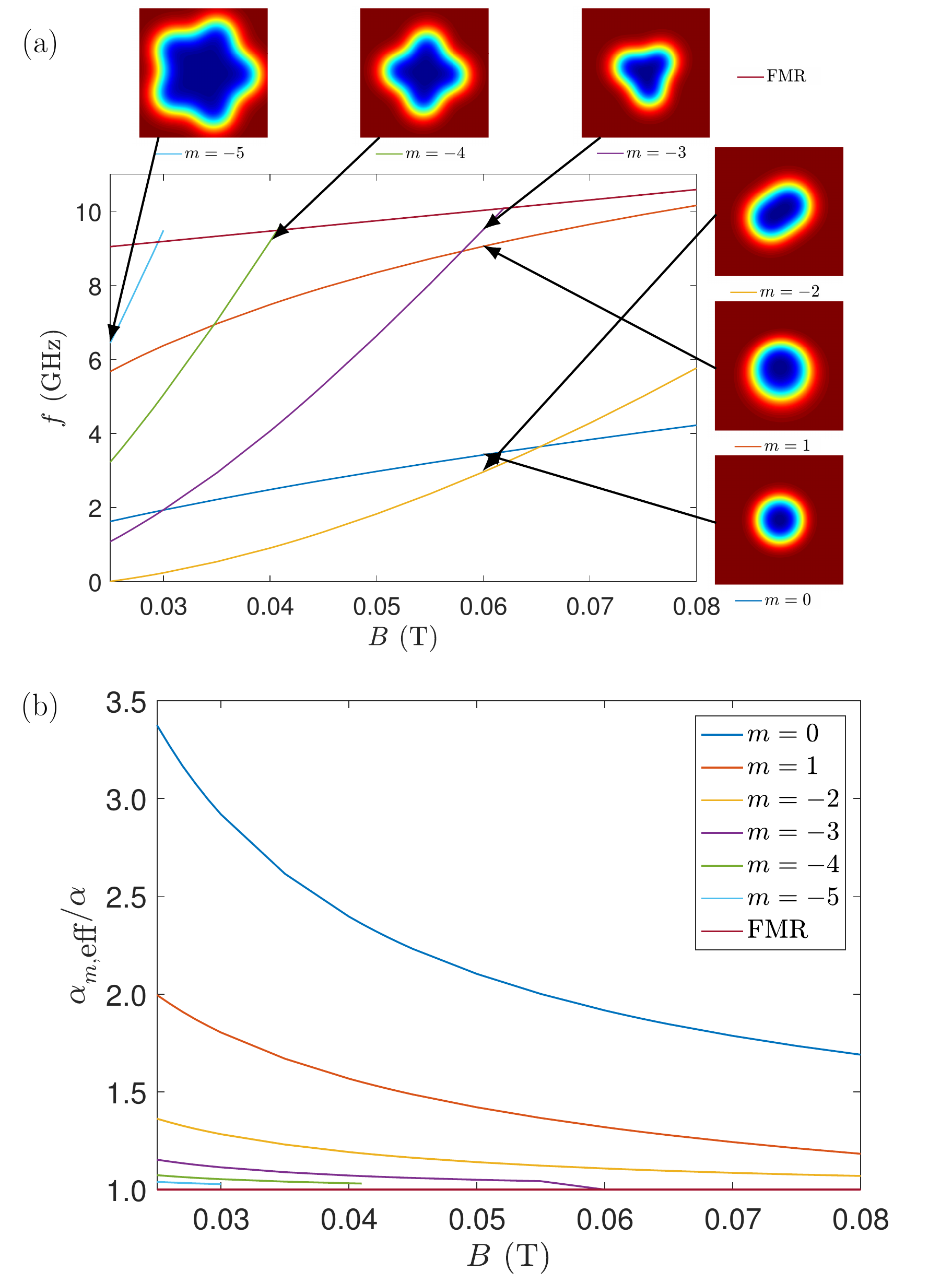}
\caption{Localized magnons in the isolated skyrmion, with the interaction parameters corresponding to the Ir|Co|Pt multilayer system from Ref.~\cite{Moreau-Luchaire}: $\mathcal{A}=10.0\,\textrm{pJ/m},\mathcal{D}=1.9\,\textrm{mJ/m}^{2},\mathcal{K}=-0.143\,\textrm{MJ/m}^{3},\mathcal{M}=0.96\,\textrm{MA/m}$. The anisotropy reflects an effective value including the dipolar interactions as a demagnetizing term, $-\mathcal{K}=-\mathcal{K}_{0}-\frac{1}{2}\mu_{0} \mathcal{M}^{2}$ with $\mathcal{K}_{0}=-0.717\,\textrm{MJ/m}^{3}$. (a) Magnon frequencies $f=\omega/2\pi$ for $\alpha=0$. Illustrations display the shapes of the excitation modes visualized as the contour plot of the out-of-plane spin components on a $1\times1\,\textrm{nm}^{2}$ grid, with red and blue colors corresponding to positive and negative $S^{z}$ values, respectively. (b) Effective damping coefficients $\alpha_{m,\textrm{eff}}$, calculated from Eq.~(6) in the main text.\label{figS2}}
\end{figure}

In order to investigate the dependence of the effective damping on the dimensionless parameters, we also performed the calculations for the parameters describing the Ir|Co|Pt multilayer system\cite{Moreau-Luchaire}. The results are summarized in Fig.~\ref{figS2}. The Ir|Co|Pt system has a larger dimensionless anisotropy value ($-\mathcal{K}^{\textrm{Ir|Co|Pt}}_{\textrm{dl}}=0.40$) than the Pd/Fe/Ir(111) system ($-\mathcal{K}^{\textrm{Pd/Fe/Ir(111)}}_{\textrm{dl}}=0.33$). Although the same localized modes are found in both cases, the frequencies belonging to the $m=0,1,-3,-4,-5$ modes in Fig.~\ref{figS2} are relatively smaller than in Fig.~2 compared to the ferromagnetic resonance frequency at the elliptic instability field where $\omega_{m=-2}=0$. This agrees with the two limiting cases discussed in the literature: it was shown in Ref.~\cite{Schutte} that for $\mathcal{K}_{\textrm{dl}}=0$ the $m=1,-4,-5$ modes are still above the ferromagnetic resonance frequency at the elliptic instability field, while in Ref.~\cite{Kravchuk} it was investigated that all modes become soft with frequencies going to zero at $\left(\mathcal{M}B\right)_{\textrm{dl}}=0$ in the point $-\mathcal{K}_{\textrm{dl}}=\frac{\pi^{2}}{16}\approx0.62$, below which a spin spiral ground state is formed in the system. Figure~\ref{figS2}(b) demonstrates that the effective damping parameters $\alpha_{m,\textrm{eff}}$ are higher at the elliptic instability field in Ir|Co|Pt than in Pd/Fe/Ir(111), showing an opposite trend compared to the frequencies.

Regarding the physical units, the stronger exchange stiffness combined with the weaker Dzyaloshinsky--Moriya interaction and anisotropy in the multilayer system leads to larger skyrmions stabilized at lower field values and displaying lower excitation frequencies. We note that demagnetization effects were only considered here as a shape anisotropy term included in $\mathcal{K}$; it is expected that this should be a relatively good approximation for the Pd/Fe/Ir(111) system with only a monolayer of magnetic material, but it was suggested recently\cite{Satywali} that the dipolar interaction can significantly influence the excitation frequencies of isolated skyrmions in magnetic multilayers.

\section{Spin dynamics simulations\label{secS5}}

For the spin dynamics simulations displayed in Fig.~3 in the main text we used an atomistic model Hamiltonian on a single-layer triangular lattice,
\begin{eqnarray}
\mathcal{H}&=&-\frac{1}{2}\sum_{\left<i,j\right>}J\boldsymbol{S}_{i}\boldsymbol{S}_{j}-\frac{1}{2}\sum_{\left<i,j\right>}\boldsymbol{D}_{ij}\left(\boldsymbol{S}_{i}\times\boldsymbol{S}_{j}\right)-\sum_{i}K\left(S^{z}_{i}\right)^{2}\nonumber
\\
&&-\sum_{i}\mu B S_{i}^{z},\label{eqnS34}
\end{eqnarray}
with the parameters $J=5.72\,\textrm{meV}$ for the Heisenberg exchange, $D=\left|\boldsymbol{D}_{ij}\right|=1.52\,\textrm{meV}$ for the Dzyaloshinsky--Moriya interaction, $K=0.4\,\textrm{meV}$ for the anisotropy, $\mu=3\mu_{\textrm{B}}$ for the magnetic moment, and $a=0.271\,\textrm{nm}$ for the lattice constant. For the transformation between the lattice and continuum parameters in the Pd/Fe/Ir(111) system see, e.g., Ref.~\cite{Hagemeister2}. The simulations were performed by numerically solving the Landau--Lifshitz--Gilbert equation on an $128\times128$ lattice with periodic boundary conditions, which was considerably larger than the equilibrium skyrmion size to minimize boundary effects. The initial configuration was determined by calculating the eigenvectors in the continuum model and discretizing it on the lattice, as shown in the insets of Fig.~2 in the main text. It was found that such a configuration was very close to the corresponding excitation mode of the lattice Hamiltonian Eq.~(\ref{eqnS34}), similarly to the agreement between the continuum and lattice equilibrium skyrmion profiles\cite{Hagemeister2}.